\documentclass[final,5p,times,twocolumn]{elsarticle}

\usepackage{amsmath}
\usepackage{amssymb}
\usepackage{graphicx}
\usepackage{bm}
\usepackage{braket}
\usepackage{enumitem}
\usepackage{tensor}
\usepackage{color}
\usepackage{combelow}

\def\omegat{\widetilde{\omega}}
\def\Et{\widetilde{E}}
\def\mytev#1{{\left<:#1:\right>_\beta}}

\journal{Physics Letters B}

\begin{document}

\begin{frontmatter}

\title{Rotating quantum states}

\author[uos]{Victor E. Ambru\cb{s}}
\ead{app10vea@sheffield.ac.uk}
\author[uos]{Elizabeth Winstanley}
\ead{e.winstanley@sheffield.ac.uk}
\address[uos]{Consortium for Fundamental Physics, School of Mathematics and Statistics, University of Sheffield,\\
Hicks Building, Hounsfield Road, Sheffield, S3 7RH, United Kingdom}

\date{\today}

\begin{abstract}
We revisit the definition of rotating thermal states for scalar and fermion fields in unbounded Minkowski space-time.
For scalar fields such states are ill-defined everywhere, but for fermion fields an appropriate definition of
the vacuum gives thermal states regular inside the speed-of-light surface.
For a massless fermion field, we derive analytic expressions for the thermal expectation values of
the fermion current and stress-energy tensor. These expressions may provide
qualitative insights into the behaviour of thermal rotating states on more complex space-time geometries.
\end{abstract}

\begin{keyword}
\PACS 03.70.+k \sep 04.62.+v
\end{keyword}

\end{frontmatter}

\section{Introduction}

In the canonical quantisation of a free field, an object of fundamental importance is the vacuum state,
from which states containing particles are constructed.
For fields of all spins, the process starts by expanding the classical field in terms of an orthonormal
basis of field modes, which are split into positive and negative frequency modes.
The expansion coefficients are promoted to operators, the expansion coefficients of the positive frequency modes
being particle annihilation operators \footnote{The adjoints of the expansion coefficients of the negative
frequency modes are also particle annihilation operators. For a real scalar field, these annihilation operators are
the same as the expansion coefficients of the positive frequency modes; for a fermion field they are different.}.
The vacuum state is defined as the state annihilated by all the particle annihilation operators.
The definition of a vacuum state is therefore dependent on how the field modes are split into
positive and negative frequency modes. This split is restricted for
a quantum scalar field by the fact that positive frequency  modes must have positive Klein-Gordon norm.
For a quantum fermion field, both positive and negative frequency fermion modes have positive Dirac norm,
so the split of the field modes into positive and negative frequency is less constrained compared with
the scalar field case. There is therefore more freedom in how the vacuum state is defined for a fermion
field, leading to more freedom in how states containing particles are defined.

In this letter we explore this difference between scalar and fermion quantum fields by considering the definition
of rotating vacuum and thermal states in Minkowski space. This toy model reveals that there are quantum states
which can be defined for a fermion field but which have no analogue for scalar fields.

\section{Rotating scalars}
We consider Minkowski space in cylindrical co-ordinates
$(t_{\text{Mink}}, \rho, \varphi_{\text{Mink}}, z)$
\footnote{Throughout this paper we use units in which $c=\hbar =k_{B}=1$.}.
We wish to define quantum states which are rigidly rotating with angular velocity $\mathbf {\Omega }$.
Choosing the $z$ axis of the coordinate system along the angular velocity
vector $\mathbf{\Omega}$, the line element of the rotating
space-time can be found by making the
transformation $\varphi = \varphi_{\text{Mink}} - \Omega t_{\text{Mink}}$, $t=t_{\text{Mink}}$ in the usual
Minkowski line element, giving:
\begin{equation}\label{eq:ds}
 ds^2 = -(1 - \rho^2\Omega^2) dt^2 + 2\rho^2 \Omega \, dt\,d\varphi + d\rho^2 +
 \rho^2 d\varphi^2 + dz^2.
\end{equation}
The Killing vector $\partial_t$, which defines the co-rotating Hamiltonian $H = i\partial_t$,  becomes null on the
speed-of-light surface (SOL), defined as the surface where $\rho = \Omega^{-1}$.
The Klein-Gordon equation for a scalar field of mass $\mu$ on the space-time \eqref{eq:ds} is:
\begin{equation}\label{eq:kg}
 \left[-(H + \Omega L_z)^2 + \frac{L_z^2}{\rho^2} + P_z^2 - \partial^2_\rho -
 \frac{\partial_\rho}{\rho} + \mu^2 \right] \Phi(x) = 0,
\end{equation}
where $P_z = -i\partial_z$ and $L_z = -i \partial_\varphi$  are the $z$ components of the momentum and
angular momentum operators, respectively. The mode solutions of \eqref{eq:kg} are:
\begin{equation}\label{eq:fkqm}
 f_{\omega k m}(x) = \frac{1}{\sqrt{8\pi^2 |\omega|}} e^{-i\omegat t + im\varphi + ikz}
 J_m(q\rho),
\end{equation}
where $J_m(q\rho)$ is the Bessel function of the first kind of order $m$,
$m$ is the eigenvalue of $L_z$, $k$ is the eigenvalue of $P_z$, $q$ is the longitudinal component of the
momentum and $\omega = \pm \sqrt{\mu^2 + q^2 + k^2}$ gives the Minkowski energy of the mode.
The eigenvalue of the Hamiltonian, $\widetilde{\omega} = \omega - \Omega m$, represents
the energy of the mode as seen by a co-rotating observer.
It is convenient to introduce the shorthand
$j = (\omega_j,k_j,m_j)$ and
\begin{equation}\label{eq:deltajjp}
 \delta(j,j') =
 \delta_{m_jm_{j'}} \delta(k_j - k_{j'})
 \frac{\delta(\omega_j-\omega_{j'})}{|\omega_j|}.
\end{equation}
Using the Klein-Gordon inner product:
\begin{equation}
 \braket{f,g} = -i\int d^3x \sqrt{-g} \, (f^* \partial^t g - g \partial^t f^*),
\end{equation}
the norm of the modes \eqref{eq:fkqm} can be calculated:
\begin{equation}
 \braket{f_j,f_{j'}} = \frac{\omega_j}{|\omega_j|} \delta(j,j').
\end{equation}
As discussed by Letaw and Pfautsch \cite{art:letaw_pfautsch}, particles must be described by modes with positive norm
($\omega_j > 0$),
implying the following expansion for the scalar field operator:
\begin{equation}\label{eq:phi}
 \Phi(x) =  \sum_{m_j = -\infty}^\infty \int_{\mu}^\infty \omega_j \, d\omega_j \int_{-p_j}^{p_j} dk_j
 \left[f_j(x) a_j+ f^*_j(x) a^\dagger_j\right],
\end{equation}
where $p_j = \sqrt{q_j^2 + k_j^2}$ is the Minkowski momentum.
The one-particle annihilation and creation operators, $a_j$ and $a^\dagger_j$, satisfy
the canonical commutation relations $[a_j, a_{j'}^\dagger] = \delta(j,j')$.
The induced vacuum state $\ket{0}$, satisfying $a_j \ket{0} = 0$,
coincides with the Minkowski vacuum \cite{art:letaw_pfautsch}.

At finite inverse temperature $\beta = T^{-1}$, Vilenkin \cite{art:vilenkin} gives the following thermal
expectation value (t.e.v.):
\begin{equation}\label{eq:aabeta}
 \braket{a_j^\dagger a_{j'}}_\beta = \frac{\delta(j,j')}{e^{\beta \widetilde{\omega_j}} -1}.
\end{equation}
The above expression cannot hold when ${\widetilde{\omega}}_{j} < 0$ \cite{art:vilenkin},
since it would imply that the vacuum expectation value of
$a_j^\dagger a_{j'}$, obtained by taking the limit $\beta\rightarrow \infty$, is non-zero, contradicting the definition
of the vacuum.
Furthermore, the
divergent behaviour of the thermal weight factor of modes with $\widetilde{\omega}$ close to $0$ renders
t.e.v.s infinite, causing rotating thermal states for scalar fields to be
ill-defined everywhere in the space-time \cite{art:vilenkin,art:duffy_ottewill}.
As discussed by \cite{art:vilenkin,art:duffy_ottewill}, a resolution to these problems is to enclose the
system inside a boundary located inside or on the SOL, restricting wavelengths such that $\omegat$ stays positive
for all values of $m$.

\section{Rotating fermions}
In the Cartesian gauge \cite{art:cota_external_symm}, a natural frame for the metric \eqref{eq:ds} can be chosen to be:
\begin{equation}\label{eq:ecart}
 e_{\hat{t}} = \partial_t - \Omega \partial_\varphi,\qquad
 e_{\hat{i}} = \partial_i.
\end{equation}
In the following, hats shall be used to indicate tensor components with respect to the tetrad, i.e.
$A^\mu = A^{\hat{\alpha}} e_{\hat{\alpha}}^\mu$.
The Dirac equation for fermions of mass $\mu$ takes the form:
\begin{equation}\label{eq:drot}
 \left[\gamma^{\hat{t}} (H + \Omega M_z) - \mathbf{\gamma}\cdot \mathbf{P} - \mu
\right] \psi(x) = 0,
\end{equation}
where the gamma matrices are in the Dirac representation \cite{book:itzykson_zuber} and the covariant derivatives
are given by:
\begin{equation}\label{eq:covd}
 i D_{\hat{t}} = H + \Omega M_z, \qquad -i D_{\hat{j}} = P_j.
\end{equation}
The momentum operators $P_j$ and angular momentum operator $M_z$ are:
\begin{equation}
 P_j = -i \partial_j,\qquad
 M_z = -i\partial_\varphi + \frac{1}{2}
 \begin{pmatrix}
  \sigma_3 & 0\\
  0 & \sigma_3
 \end{pmatrix}.
\end{equation}

The Dirac equation \eqref{eq:drot} admits the following solutions:
\begin{equation}\label{eq:U}
 U_{Ekm}^\lambda(x) = \frac{1}{\sqrt{8\pi^2}} e^{-i\widetilde{E} t + ik z}
 \begin{pmatrix}
  \sqrt{1 + \frac{\mu}{E}} \, \phi^\lambda_{Ekm} \\
  \frac{2\lambda E}{|E|} \sqrt{1 - \frac{\mu}{E}} \, \phi^\lambda_{Ekm}
 \end{pmatrix},
\end{equation}
where the two-spinor $\phi^\lambda_{Ekm}$ is defined as:
\begin{equation}
 \phi^\lambda_{Ekm}(\rho, \varphi) = \frac{1}{\sqrt{2}}
 \begin{pmatrix}
  \sqrt{1 + \frac{2\lambda k}{p}} e^{im\varphi} J_m(q\rho)\\
  2i\lambda \sqrt{1 - \frac{2\lambda k}{p}} e^{i(m+1)\varphi} J_{m+1}(q\rho)
 \end{pmatrix},
\end{equation}
where $\lambda$ is the helicity \cite{art:cota_external_symm,book:itzykson_zuber},
$p = \sqrt{q^2 + k^2}$ is the magnitude of the momentum and
$E = \pm \sqrt{p^2 + \mu^2}$ controls the sign of the Minkowski energy of the
mode. The eigenvalues of the Hamiltonian are $\widetilde{E} = E - \Omega (m + \tfrac{1}{2})$, representing,
as in the scalar case, the energy seen by a co-rotating observer.
The notations $j = (E_j,k_j,m_j,\lambda_j)$ and
\begin{equation}
 \delta(j,j') =
 \delta_{\lambda_j\lambda_{j'}} \delta_{m_jm_{j'}} \delta(k_j - k_{j'})
 \frac{\delta(E_j-E_{j'})}{|E_j|}
\end{equation}
are useful to refer to modes and their norms. The latter can be computed using the Dirac inner product:
\begin{equation}
 \braket{\psi,\chi} = \int d^3x \sqrt{-g} \, \psi^\dagger(x) \chi(x).
\end{equation}
It can be shown that $\braket{U_j, U_{j'}} = \delta(j,j')$ for all possible labels $j$, $j'$.
After choosing a suitable definition for
particle modes (i.e.~a range for the labels in $j$), the anti-particle modes can be constructed using charge
conjugation \cite{art:cota_external_symm,book:itzykson_zuber}: $V_j = i \gamma^{\hat{2}} U_j^*$.
Hence, $V_j$ automatically inherits
the same normalisation as $U_j$, namely: $\braket{V_j,V_{j'}} = \delta(j,j')$.
Therefore there is no restriction on how the split into particle and anti-particle modes is performed,
as long as the charge conjugation symmetry is preserved.

According to Vilenkin \cite{art:vilenkin}, the definition of particles
for co-rotating observers should be the same as for inertial Minkowski observers, with the field operator written as:
\begin{multline}\label{eq:psi_V}
 \psi_V(x) = \sum_{\lambda_j = \pm \frac{1}{2}}
 \sum_{m_j = -\infty}^{\infty}
 \int_{\mu}^\infty E_j \, dE_j \int_{-p_j}^{p_j} dk_j\\
 \times \left[U_j(x) b_{j;V} + V_j(x) d^\dagger_{j;V}\right].
\end{multline}
Vilenkin's quantisation is equivalent to the one suggested by
Letaw and Pfautsch \cite{art:letaw_pfautsch} for the scalar field, yielding a vacuum state equivalent to the
Minkowski vacuum. In contrast, Iyer \cite{art:iyer} argues that the modes which represent particles
for a co-rotating observer have positive frequency with respect to the co-rotating Hamiltonian, implying the following
expression for the field operator:
\begin{multline}\label{eq:psi_I}
 \psi_I(x) = \sum_{\lambda_j = \pm \frac{1}{2}}
 \sum_{m_j = -\infty}^{\infty}
 \int_{\Et _{j} > 0, |E_{j}| > \mu}^\infty E_j \, dE_j \int_{-p_j}^{p_j} dk_j\\
 \times \left[U_j b_{j;I} + V_j(x) d^\dagger_{j;I}\right],
\end{multline}
with the integral with respect to $E_j$ running over both positive and negative values of $E_j$, as long as
$\widetilde{E}_j  > 0$ and $|E_j| > \mu$. Both quantisation methods lead to the canonical anti-commutation relations
 $\{b_j, b_{j'}^\dagger\} = \{d_j, d_{j'}^\dagger\} = \delta(j,j')$.
The ensuing quantum field theory differs in the two pictures, as the vacuum state corresponding to Iyer's quantisation
differs from the Minkowski vacuum. This can be seen by looking at the connection between the Iyer and Vilenkin one-particle
operators:
\begin{equation}\label{eq:psi_iyer}
 b_{j;I} =
 \begin{cases}
  b_{j;V} & \text{$E_j$ is positive},\\
  i^{2m+1} d^\dagger_{\overline{\jmath};V} & \text{$E_j$ is negative},
 \end{cases}
\end{equation}
and similarly for $d_{j;I}$, where $\overline{\jmath} = (-E_j,-k_j,-m_j-1,\lambda_j)$.
Thus, the Vilenkin vacuum state (i.e.~the non-rotating Minkowski vacu\-um) contains particles
as defined according to Iyer's quantisation.
Similarly, the Iyer vacuum contains particles as defined according to Vilenkin's quantisation (i.e.~relative to the
Minkowski vacuum).

Vilenkin \cite{art:vilenkin} also considered rotating thermal states for fermions.
In analogy with (\ref{eq:aabeta}), he gives the following t.e.v.s relative to the
Minkowski vacuum \cite{art:vilenkin}:
\begin{equation}\label{eq:bbddbeta}
 \braket{b_j^\dagger b_{j'}}_\beta = \braket{d_j^{\dagger} d_{j'}}_\beta
 =\frac{\delta(j,j')}{e^{\beta \widetilde{E_j}}+ 1}.
\end{equation}
As in the scalar case, \eqref{eq:bbddbeta} is not valid when $\widetilde{E}_j < 0$ \cite{art:vilenkin}.
However, in contrast to the scalar case, the modes with negative $\widetilde{E}_j$
can be eliminated from the set of particle modes by using Iyer's quantisation, without enclosing the system within a boundary inside the SOL.
Furthermore, unlike the thermal factor for scalars \eqref{eq:aabeta},
the Fermi-Dirac density of states factor \eqref{eq:bbddbeta} is regular for all $\widetilde{E}_j$.

Eq.~\eqref{eq:bbddbeta} can be used to construct the t.e.v.s of the neutrino charge current operator and of
the SET:
\begin{subequations}\label{eq:ops}
\begin{align}
 J^{\hat{\alpha}}(x) =& \frac{1}{2}
 \left[\overline{\psi}, \gamma^{\hat{\alpha}} \frac{1+\gamma^{\hat{5}}}{2}\psi\right],\\
 T_{\hat{\alpha}\hat{\sigma}}(x) =& -\frac{i}{4} \left\{
 [\,\overline{\psi},\gamma_{(\hat{\alpha}} D_{\hat{\sigma})} \psi\,] -
 [\,\overline{D_{(\hat{\alpha}}\psi} \gamma_{\hat{\sigma})}, \psi\,]\right\}.
\end{align}
\end{subequations}
Using the Vilenkin quantisation, we find the following t.e.v.s relative to the Minkowski vacuum:
\begin{subequations}\label{eq:tevs_comp}
\begin{align}
 \mytev{\left[\overline{\psi}\psi\right]_V} =& -\mu S_{000}^+,\label{eq:tevs_comp_ppsi}\\
 \mytev{J^{\hat{z}}_V} =& -\tfrac{1}{2} S_{100}^-,\\
 \mytev{T_{V;\hat{t}\hat{t}}} =& S_{200}^+,\\
 \mytev{T_{V;\hat{\rho}\hat{\rho}}} =& S_{020}^+ - \rho^{-1} S_{011}^{\times},\\
 \mytev{T_{V;\hat{\varphi}\hat{\varphi}}} =& \rho^{-1} S_{011}^{\times},\\
 \mytev{T_{V;\hat{z}\hat{z}}} =& S_{200}^+ - S_{020}^+ - \mu^2 S_{000}^+,\\
 \mytev{T_{V;\hat{t}\hat{\varphi}}} =& \tfrac{1}{4} \rho^{-1} S_{100}^- - \tfrac{1}{2} \rho^{-1} S_{101}^+ -
 \tfrac{1}{2} S_{110}^{\times},
\end{align}
\end{subequations}
where the fermion condensate $\mytev{\left[\overline{\psi}\psi\right]_V}$ vanishes when $\mu = 0$.
The functions $S^\pm_{abc}$ and $S^\times_{abc}$ introduced above are defined as:
\begin{equation}\label{eq:bblocks}
 S^*_{abc} = \frac{1}{\pi^2} \sum_{m = -\infty}^\infty \int_\mu^\infty \frac{dE}{1 + e^{\beta \widetilde{E}}}
 \int_0^p dk\, E^a q^b \left( m + \tfrac{1}{2} \right)^c J^*_m(q\rho),
\end{equation}
where $* \in \{+,-,\times\}$ and the functions $J^*_m$ are given by:
\begin{equation}\label{eq:Jcombs}
 J^\pm_m(z) \equiv J_m^2(z) \pm J_{m+1}^2(z), \qquad
 J^{\times}_m(z) = 2 J_m(z) J_{m+1}(z).
\end{equation}

Except when $\mu = 0$, numerical integration must be used to calculate $S^*_{abc}$ for arbitrary values
of the mass. For massless fermions, the method outlined in \ref{app:bblocks}
can be followed to obtain the following exact results ($\varepsilon = 1 - \rho^2 \Omega^2$):
\begin{subequations}\label{eq:tevs}
\begin{align}
 \frac{1}{\mu} \mytev{\left[\overline{\psi}\psi\right]_V} =& -\frac{1}{6\beta^2 \varepsilon} -
 \frac{\Omega^2}{8\pi^2 \varepsilon^2} \left(\frac{2}{3} + \frac{\varepsilon}{3}\right),\label{eq:ppsi}\\
 \mytev{J^{z}_V} =& -\frac{\Omega}{12\beta^2\varepsilon^2} -
 \frac{\Omega^3}{48\pi^2\varepsilon^3}\left(4 - 3\varepsilon\right).\label{eq:jz}
\end{align}
To evaluate the massless limit of the t.e.v. of the fermion condensate $\mytev{\left[\overline{\psi}\psi\right]_V}$,
the latter was divided by the mass factor in Eq.~\eqref{eq:tevs_comp_ppsi}.
In the above, the hat has been dropped from the index of $J^z$ to indicate that the result is with respect
to the coordinate basis. For the $z$ component, this coincides with the tetrad component.
The t.e.v. of the SET with respect to the coordinate basis has the following components:
\begin{align}
 \left<:T_{V;tt}:\right>_\beta =& \frac{7\pi^2}{60\beta^4\varepsilon} + \frac{\Omega^2}{8\beta^2\varepsilon^2}
 \left(\frac{4}{3} - \frac{1}{3}\varepsilon\right)\nonumber\\
 &+ \frac{\Omega^4}{64\pi^2\varepsilon^3}\left(\frac{8}{9} + \frac{56}{45}\varepsilon -
 \frac{17}{15}\varepsilon^2\right),\label{eq:Ttt}\\
 \left<:T_{V;\rho\rho}:\right>_\beta =& \frac{7\pi^2}{180\beta^4 \varepsilon^2} +
 \frac{\Omega^2}{24\beta^2\varepsilon^3} \left(\frac{4}{3} - \frac{1}{3} \varepsilon\right)\nonumber\\
 &+ \frac{\Omega^4}{192\pi^2\varepsilon^4}\left(8-\frac{88}{15}\varepsilon - \frac{17}{15}\varepsilon^2\right),
 \label{eq:Trr}\\
 \frac{1}{\rho} \left<:T_{V;\varphi t}:\right>_\beta =& -\rho\Omega \Bigg\{\frac{7\pi^2}{60\beta^4\varepsilon^2}
 + \frac{13\Omega^2}{72\beta^2\varepsilon^3}\left(\frac{16}{13} - \frac{3}{13}\varepsilon\right)\nonumber\\
 &+\frac{119\Omega^4}{960\pi^2\varepsilon^4}\left(\frac{200}{119}- \frac{64}{119}\varepsilon -
 \frac{1}{7}\varepsilon^2\right)\Bigg\},\label{eq:Ttp}\\
 \frac{1}{\rho^2} \left<:T_{V;\varphi\varphi}:\right>_\beta =&
 \frac{7\pi^2}{180\beta^4\varepsilon^3} \left(4 - 3\varepsilon\right)
 + \frac{\Omega^2}{24\beta^2\varepsilon^4}\left(8 - 8\varepsilon + \varepsilon^2\right)\nonumber\\
 &+ \frac{\Omega^4}{192\pi^2\varepsilon^5}\left(64 - \frac{456}{5}\varepsilon + \frac{124}{5}\varepsilon^2 +
 \frac{17}{5}\varepsilon^3\right),\label{eq:Tpp}
\end{align}
and $\mytev{T_{V;zz}} = \mytev{T_{V;\rho\rho}}$ for any value of the mass $\mu$.
\end{subequations}
The connection between the tetrad and coordinate basis components is made through:
\begin{subequations}\label{eq:Tcoords}
\begin{align}
 T_{tt} =& T_{\hat{t}\hat{t}} + 2\rho \Omega T_{\hat{t}\hat{\varphi}} + \rho^2 \Omega^2 T_{\hat{\varphi}\hat{\varphi}},\\
 T_{t\varphi} =& \rho T_{\hat{t}\hat{\varphi}} + \rho^2 \Omega T_{\hat{\varphi}\hat{\varphi}},\\
 T_{\varphi\varphi} =& \rho^2 T_{\hat{\varphi}\hat{\varphi}}.
\end{align}
\end{subequations}

\begin{figure*}
\caption{Plots of the t.e.v.s in Eqs.~\eqref{eq:tevs_comp} using Iyer's quantisation for fermions
with mass $\mu = 0$ (thin lines) and $\mu = 2\Omega$, at inverse temperatures (from top to bottom)
$0.8$, $1.0$, $1.25$ and $2.0$, in units of $\Omega^{-1}$.
(a) $\mytev{T_{I;tt}}$ against $\rho \Omega$,
(b)-(g) Log-log plots of Eqs.~\eqref{eq:tevs_comp}, showing the polynomial nature of the divergence as the
SOL is approached. It can be seen that the massive t.e.v.s diverge at exactly the same rate as in the
massless case.}
\label{fig}
\begin{tabular}{cc}
\includegraphics[width=0.95\columnwidth]{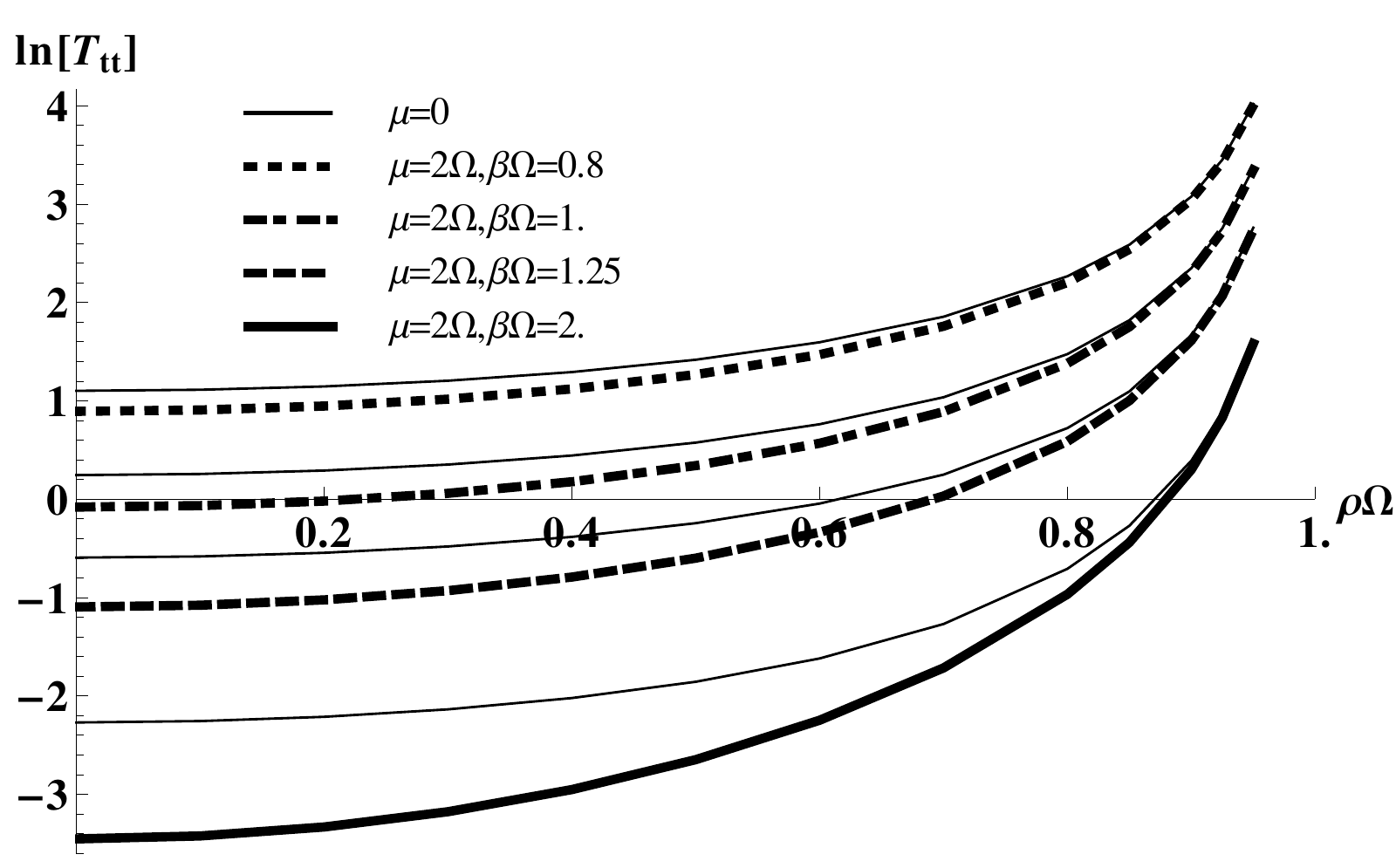} &
\includegraphics[width=0.95\columnwidth]{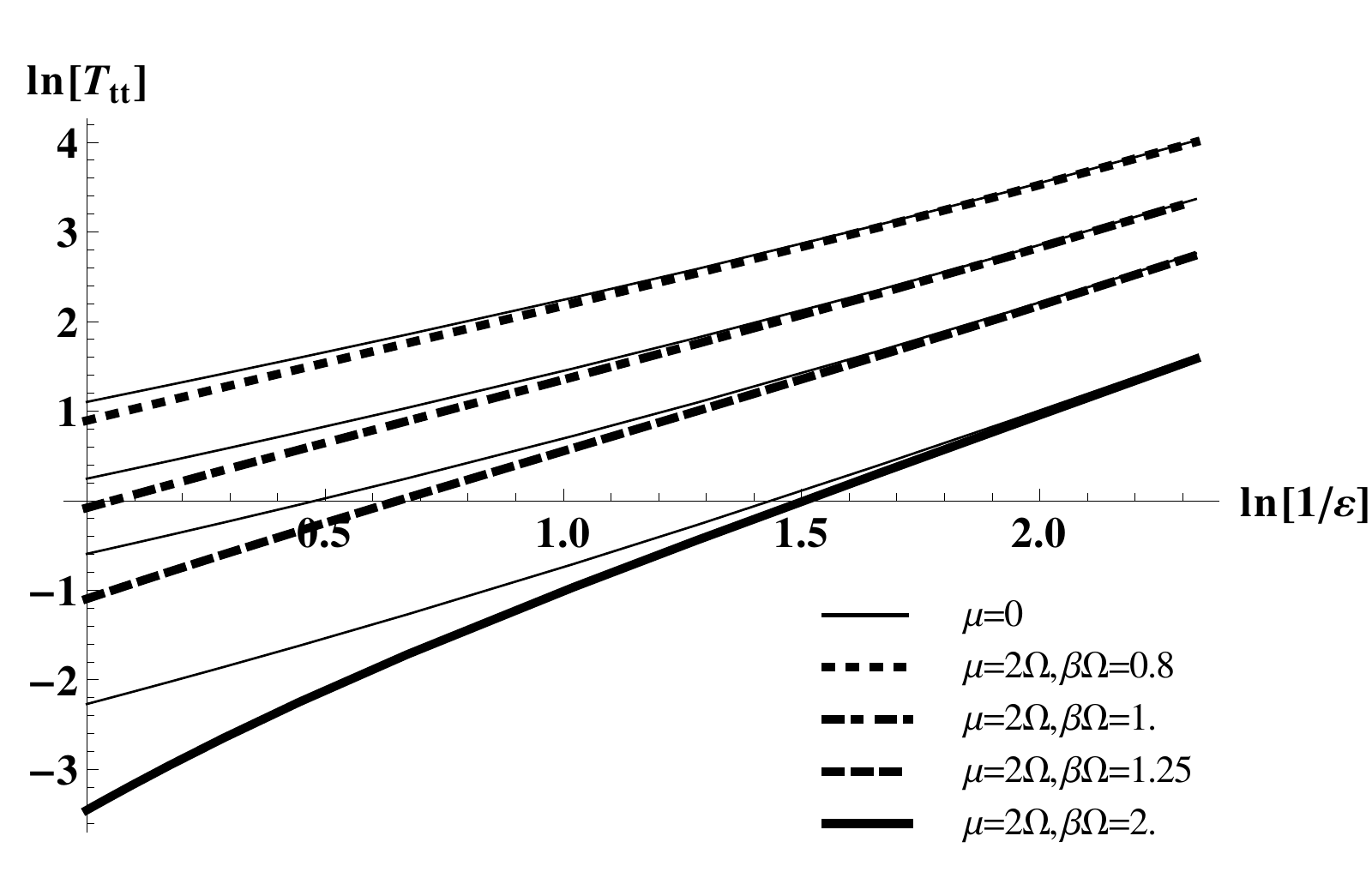} \\
\includegraphics[width=0.95\columnwidth]{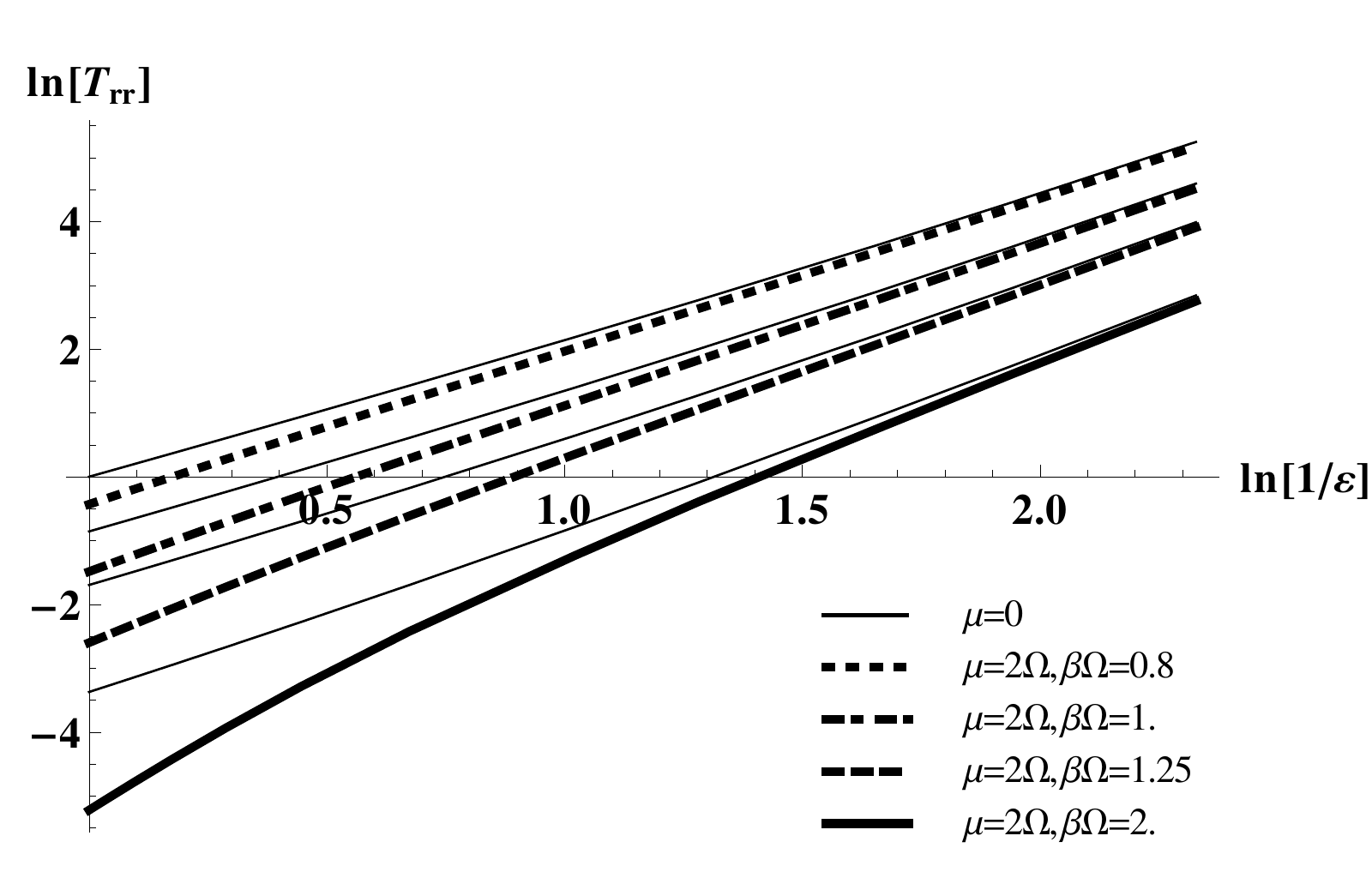} &
\includegraphics[width=0.95\columnwidth]{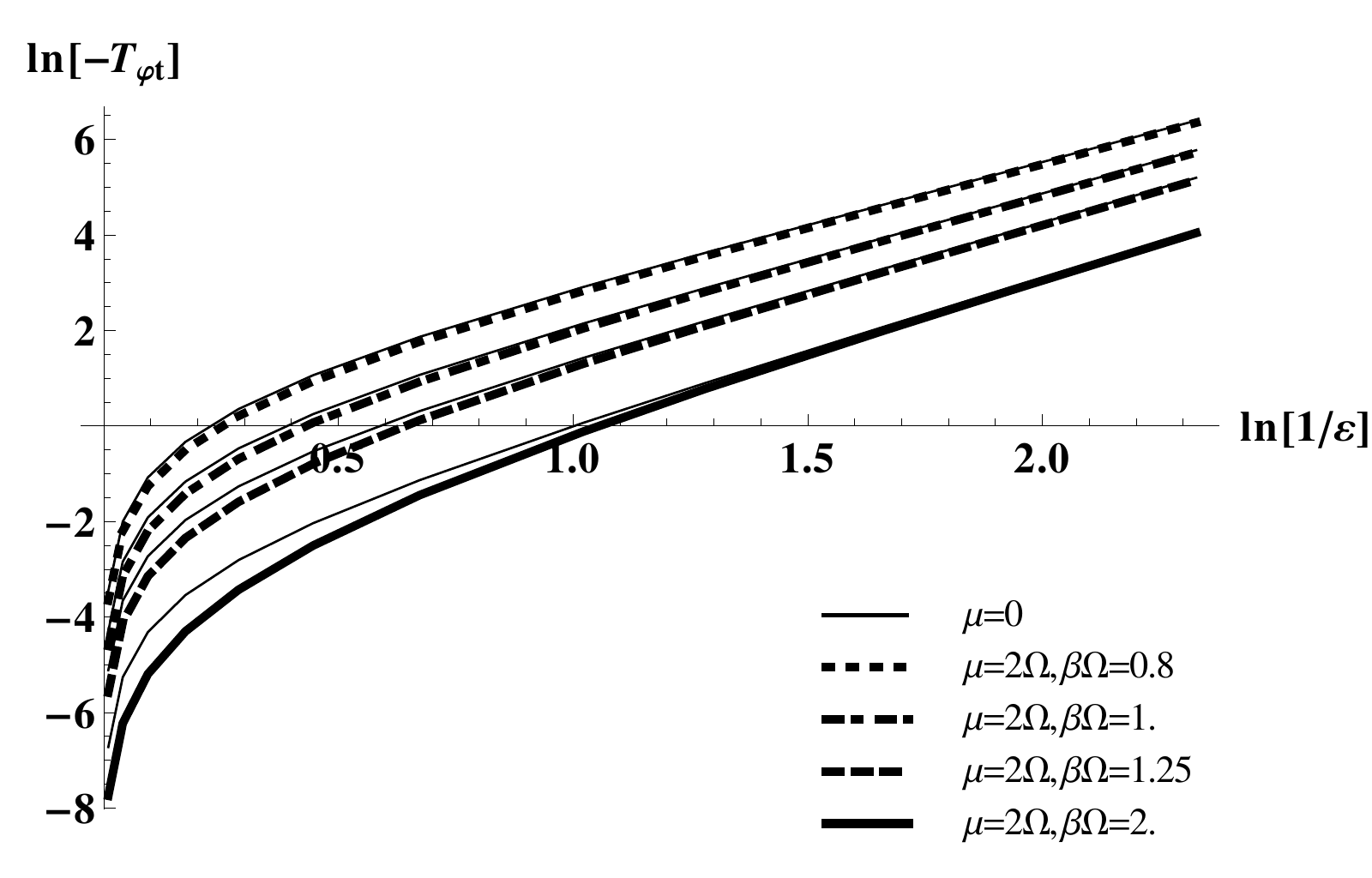} \\
\includegraphics[width=0.95\columnwidth]{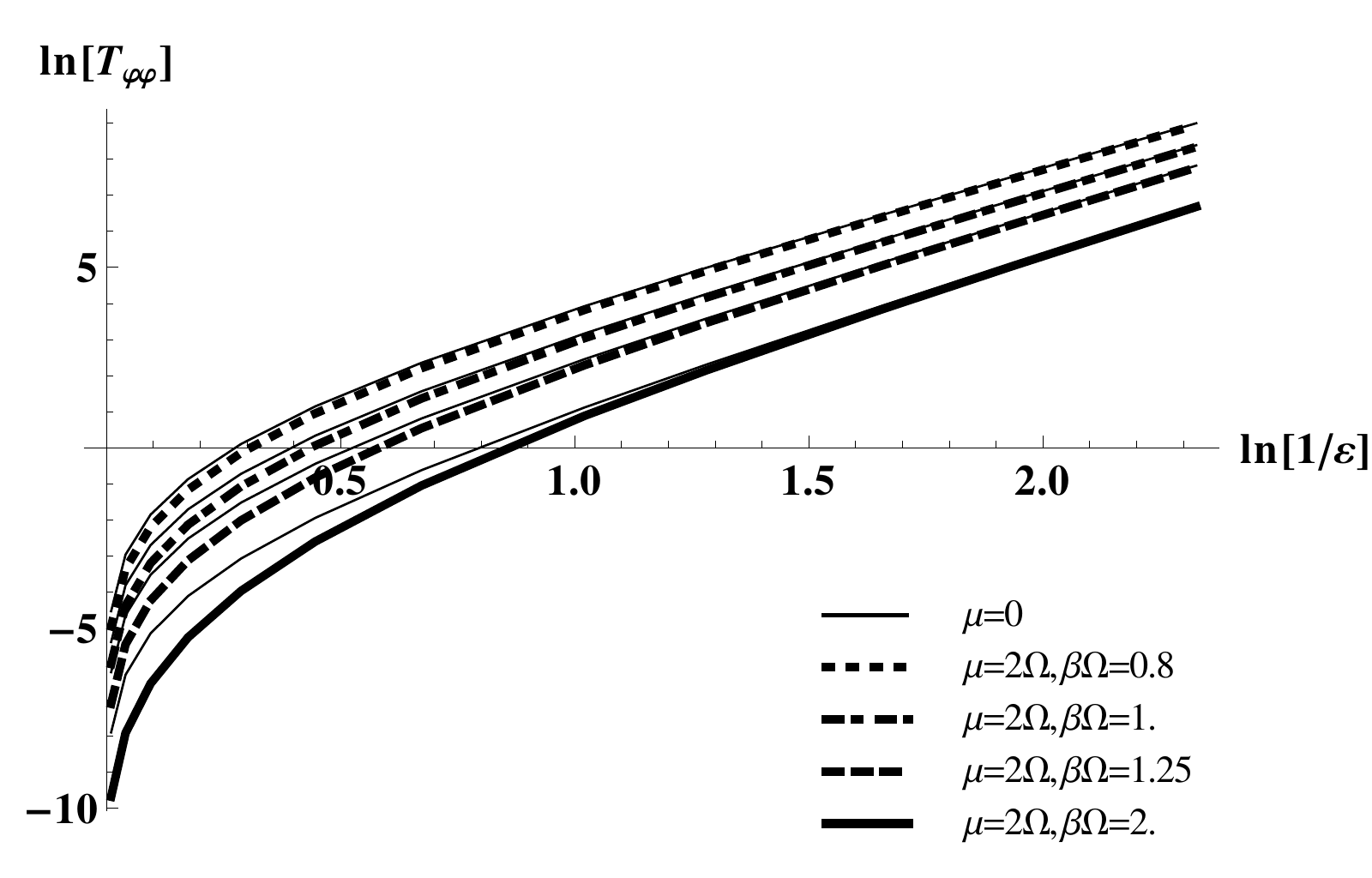} &
\includegraphics[width=0.95\columnwidth]{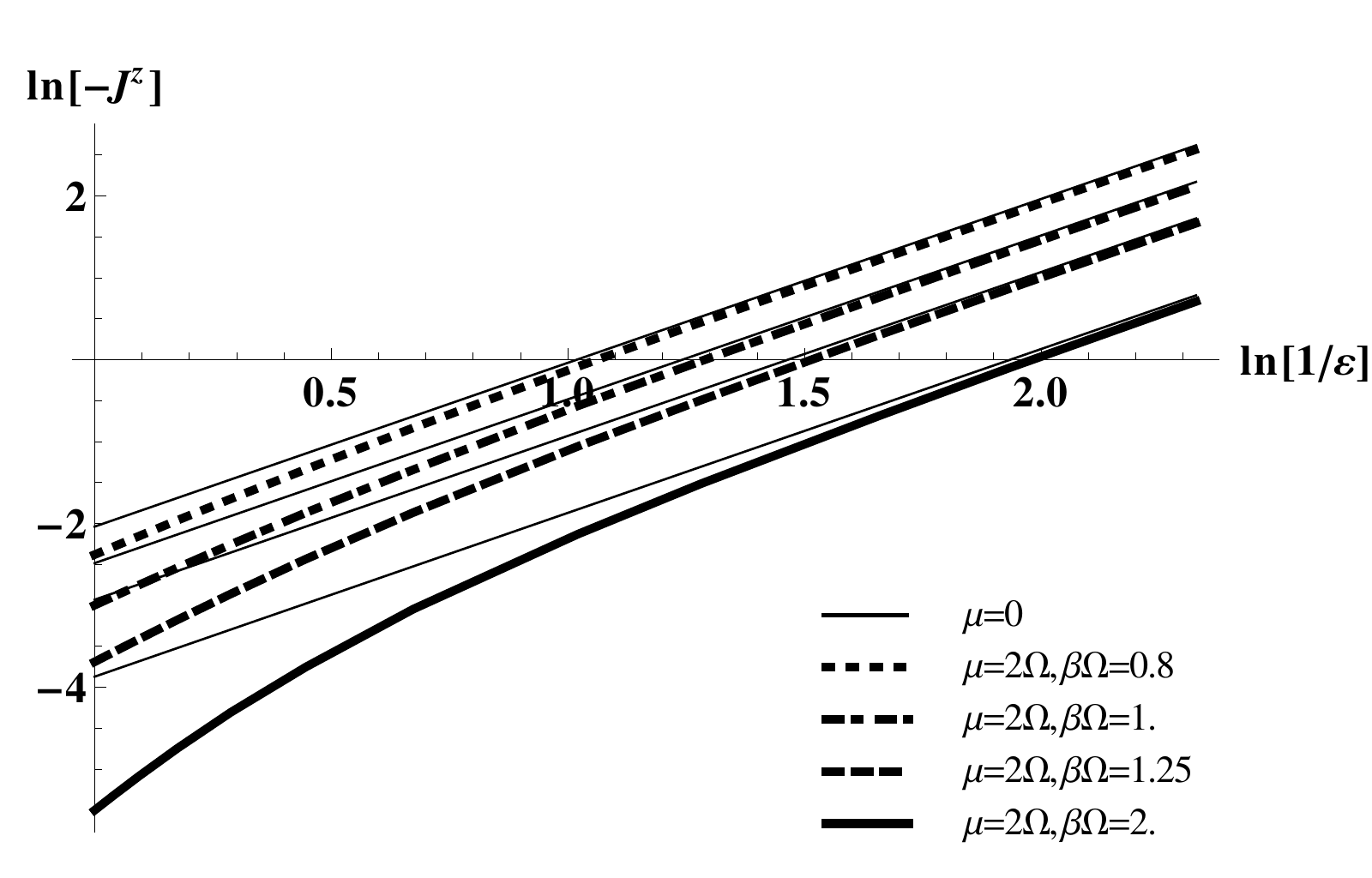} \\
\multicolumn{2}{c}{\includegraphics[width=0.95\columnwidth]{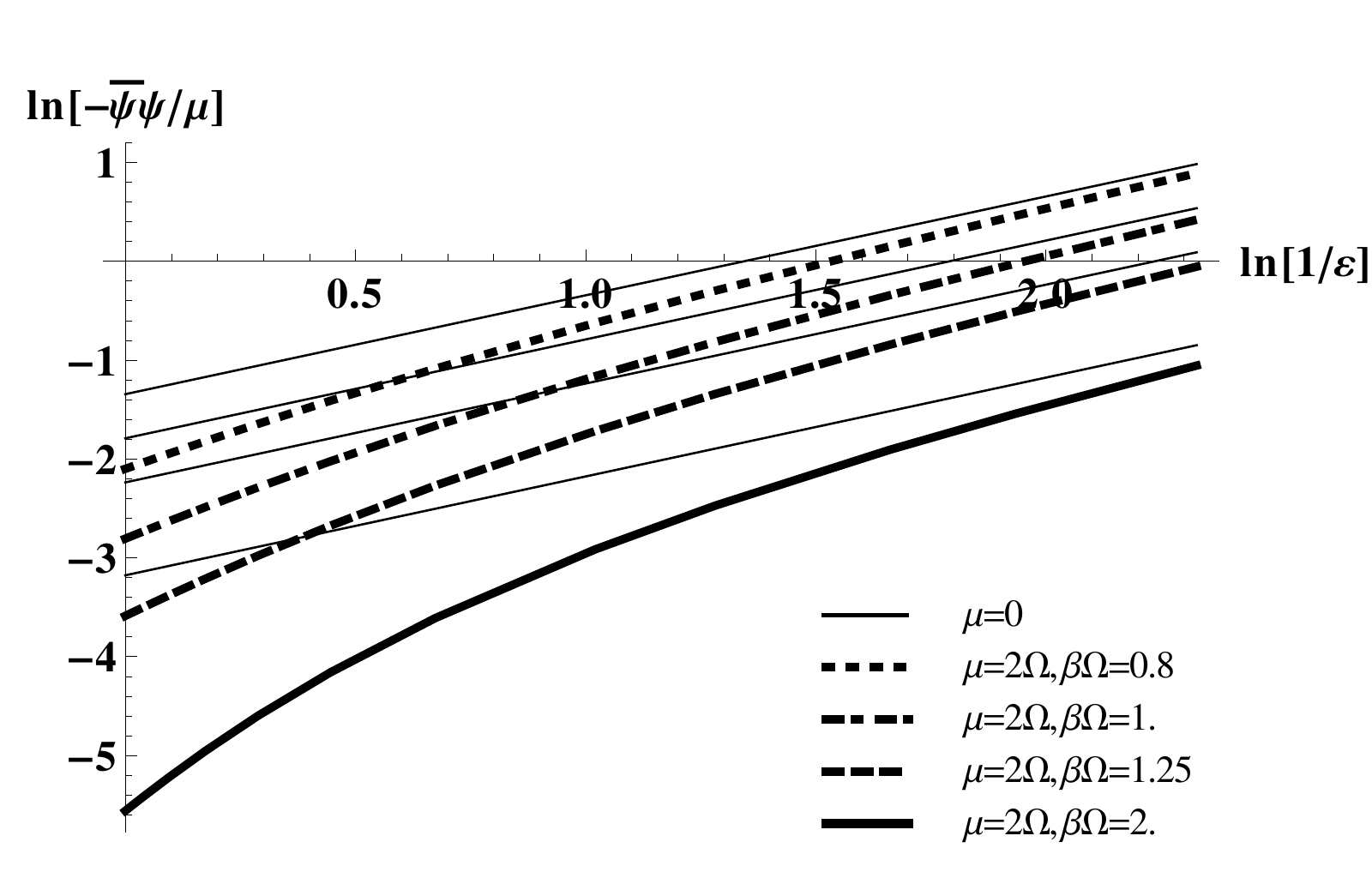}}
\end{tabular}
\end{figure*}

The analytic results for $\mytev{\left[\overline{\psi}\psi\right]_V}$, $\mytev{J^{\hat{z}}_V}$ and
$\left<:T_{V;tt}:\right>_\beta$ given by Eqs.~\eqref{eq:tevs} reveal a number of physical features.
Firstly, they all contain contributions which are independent of the temperature (equivalently, independent of $\beta$).
These terms are unphysical as the t.e.v.s should vanish when the temperature is set to zero ($\beta \rightarrow \infty$).
These temperature-independent terms are generated by modes with $\widetilde{E} < 0$ and arise because the Vilenkin
quantisation \cite{art:vilenkin} has been used.
If the Iyer quantisation \cite{art:iyer} is employed, analytic expressions for the corresponding t.e.v.s for massless
fermions
$\mytev{\left[\overline{\psi}\psi\right]_I}$,
$\left<:J_{I}^{z}:\right>_\beta$ and
$\left<:T_{I;\alpha \sigma }:\right>_\beta$ are found from \eqref{eq:tevs}
by subtracting the temperature-independent parts,
for example,
\begin{equation}
\left<:J_{I}^{z}:\right>_\beta =-\frac{\Omega}{12\beta^2\varepsilon^2}.
\end{equation}
The temperature-independent contributions to Eqs.~\eqref{eq:tevs} are the expectation
values of the Iyer vacuum relative to the Vilenkin (Minkowski) vacuum. Since these are non-zero and depend on $\rho$,
we see that the Iyer vacuum is not equivalent
to, and has fewer symmetries than, the maximally symmetric Vilenkin vacuum.
The temperature-independent contributions to Eqs.~\eqref{eq:tevs} are absent from the t.e.v.s calculated using the Iyer quantisation.

If $\Omega = 0$, there is no rotation and the t.e.v.s \eqref{eq:tevs} reduce to the usual Minkowski t.e.v.s.
If $\Omega \neq 0$, on the axis of rotation we have $\varepsilon =1$ and the expressions in round brackets in \eqref{eq:tevs}
evaluate to unity.
In this case the t.e.v.s take the form of the Minkowski values plus an
$\Omega$-dependent correction.
A similar effect is found for rotating thermal states for a scalar field inside a
reflecting cylinder \cite{art:duffy_ottewill}.

The t.e.v.s \eqref{eq:tevs} are finite as long as $\varepsilon > 0$ but diverge as
$\varepsilon \rightarrow 0$ and the SOL
is approached.
Using the Iyer quantisation (when the $\beta $-independent terms are absent),
the fermion condensate $\mytev{\left[\overline{\psi}\psi\right]_I}$ diverges as $\varepsilon^{-1}$,
the current $\left<:J_{I}^{z}:\right>_\beta$ diverges as $\varepsilon ^{-2}$ and
the SET components $\left<:T\indices{_I^\alpha_\sigma}:\right>_\beta $ diverge as
$\varepsilon^{-4}$.

For the case of arbitrary mass, numerical methods can be employed.
In Figure~\ref{fig} we show the t.e.v.s in Eqs.~\eqref{eq:tevs_comp} using the Iyer quantisation for massive fermions.
As expected, increasing the mass damps the t.e.v.s. The damping becomes more pronounced as the temperature is
decreased ($\beta$ is increased). It can be seen from the log-log plots that as the SOL is approached,
the t.e.v.s in the massive case diverge at exactly the same rate as for massless fermions.
The mass contributes
corrections which diverge at subleading orders, so that, close to the SOL, the quanta behave as if they were massless.

\section{Conclusions}

We have studied the construction of rotating states for scalar and fermion fields in four-dimensional Minkowski space.
Our analysis has demonstrated that the definition of fermion quantum states is
less constrained than for scalar fields.
This is due to more freedom in how the split into particle and anti-particle modes is performed for
fermion fields, since {\em {all}} fermion modes
have positive norm (the Dirac norm is positive definite but the Klein-Gordon norm is not).
We have considered two possible quantisation schemes \cite{art:vilenkin,art:iyer} for rotating fermion states,
which yield two inequivalent vacuum states.
We have computed thermal expectation values (t.e.v.s) for rotating states relative to these two vacua.
In the Vilenkin scheme \cite{art:vilenkin}, the vacuum state is equivalent to the (non-rotating) Minkowski vacuum, and t.e.v.s
for rotating states relative to this vacuum include temperature-independent terms which arise from low energy modes having positive frequency as seen by a non-rotating observer but negative frequency as seen by a rotating observer.
As discussed by Vilenkin \cite{art:vilenkin}, these low energy modes (and hence the temperature-independent terms in the t.e.v.s) can be removed by enclosing the system within a boundary inside the speed-of-light surface \footnote{Enclosing relativistic fermions within a boundary presents some difficulties \cite{art:vilenkin}. We will return to this issue
in a future publication \cite{art:aw}.}.
Using the Iyer scheme \cite{art:iyer}, the temperature-independent terms in the t.e.v.s are absent, without enclosing the system in a boundary or
otherwise modifying the particle spectrum.
As discussed in \cite{art:iyer}, we emphasize that the Iyer vacuum state cannot be defined for a quantum scalar field (which is restricted to
the Minkowski vacuum).
In addition, while rotating thermal states for scalar fields are ill-defined everywhere on the unbounded space-time \cite{art:duffy_ottewill},
we have constructed fermion rotating thermal states which are regular inside the speed-of-light surface.

In this paper we have considered the toy model of rotating states in Minkowski space-time.
However, the main physical features extend to curved space-times.
For example, recent work on the construction of quantum states for fermion fields on a rotating
Kerr black hole \cite{art:kermions}
has also demonstrated the existence of fermion states which have no analogue for scalar fields.
Furthermore, the simplicity of our toy model has enabled us to derive analytic expressions for the
thermal expectation values for massless fermions, which could provide
qualitative insights into the behaviour of thermal rotating states on more complex space-time geometries,
for example, the nature of the divergence as the speed-of-light surface is approached.

\section*{Acknowledgements}
This work is supported by the Lancaster-Manchester-Sheffield Consortium for
Fundamental Physics under STFC grant ST/J000418/1,
the School of Mathematics and Statistics at the University of Sheffield
and the European Cooperation in Science and Technology (COST) action MP0905
``Black Holes in a Violent Universe''.

\appendix

\section{Fermi-Dirac integrals for massless rotating states}\label{app:bblocks}
To compute the functions $S^*_{abc}$, defined in Eq.~\eqref{eq:bblocks}, the Fermi-Dirac density of states
factor can be expanded about $\Omega = 0$:
\begin{equation}
 \frac{1}{1 + e^{\beta[E - \Omega(m + \frac{1}{2})]}} =
 \sum_{n = 0}^\infty \frac{(-\Omega)^n}{n!} \left( m + \tfrac{1}{2} \right)^n\,
 \frac{d^n}{dE^n} \left(\frac{1}{1 + e^{\beta E}}\right),
\end{equation}
leading to:
\begin{multline}\label{eq:bblocks_aux}
 S^*_{abc} = \frac{1}{\pi^2} \sum_{n = 0}^\infty \frac{(-\Omega)^n}{n!}
 \int_\mu^\infty dE\, E^a \frac{d^n}{dE^n} \left(\frac{1}{e^{\beta E} + 1}\right)\\
 \times \int_0^p dk\, q^b
 \sum_{m = -\infty}^\infty \left( m + \tfrac{1}{2} \right)^{n+c} J^*_m(q\rho).
\end{multline}

\paragraph{Sum over $m$}
The sum over $m$ in Eq.~\eqref{eq:bblocks_aux} vanishes unless
$n + c$ is even for $* = +$ and odd for $* \in \{-, \times\}$.
To perform the sum, the following formula can be used to rewrite the product
of two Bessel functions as an infinite sum:
\begin{multline}
 J_\nu(z) J_\mu(z) = \sum_{k = 0}^\infty \frac{(-1)^k}{k! \Gamma(\nu + k + 1) \Gamma(\mu + k + 1)}\\
 \times \frac{\Gamma(\nu + \mu + 2k + 1)}{\Gamma(\nu + \mu + k + 1)} \left(\frac{z}{2}\right)^{2k + \nu + \mu}.
\end{multline}
After the sum over $m$ is performed, the above series terminates after a finite number of terms, as follows:
\begin{subequations}\label{eq:Jsums}
\begin{align}
 \sum_{m = -\infty}^\infty \left( m + \tfrac{1}{2} \right)^{2n} J^+_m(z) =&
 \sum_{j = 0}^n \frac{2\Gamma(j + \tfrac{1}{2})}{j! \sqrt{\pi}} s^+_{n,j} z^{2j},\label{eq:Jsum}\\
 \sum_{m = -\infty}^\infty \left( m + \tfrac{1}{2} \right)^{2n + 1} J^-_m(z) =&
 \sum_{j = 0}^n \frac{2\Gamma(j + \tfrac{1}{2})}{j! \sqrt{\pi}} s^-_{n,j} z^{2j},\label{eq:Jmsum}\\
 \sum_{m = -\infty}^\infty \left( m + \tfrac{1}{2} \right)^{2n + 1} J^{\times}_m(z) =&
 \sum_{j = 0}^n \frac{2\Gamma(j + \tfrac{1}{2})}{j! \sqrt{\pi}} s^\times_{n,j} z^{2j+1},\label{eq:Jtsum}
\end{align}
\end{subequations}
where $s^+_{n,j}$ can be shown to equal:
\begin{equation}\label{eq:sp}
 s^+_{n,j} = \frac{1}{(2j+1)!} \lim_{\alpha \rightarrow 0} \frac{d^{2n+1}}{d\alpha^{2n+1}}
 \left(2 \sinh \tfrac{\alpha}{2}\right)^{2j + 1}.
\end{equation}
It is clear that $s^+_{n,j}$ vanishes when $j > n$.
Using the following properties of $J^*_m$:
\begin{equation}
 \frac{d}{dz} \left( z J^+_m(z) \right) = (2m+1) J^-_{m}(z),\qquad
 \frac{d}{dz} \left( z J^\times_m(z) \right) = 2z J^-_m(z),
\end{equation}
it can be shown that $s^-_{n,j}$ and $s^\times_{n,j}$ are related to $s^+_{n,j}$ through:
\begin{equation}
 s^-_{n,j} = \left( j + \tfrac{1}{2} \right) s^+_{n,j}, \qquad
 s^\times_{n,j} = \frac{j + \tfrac{1}{2}}{j + 1} s^+_{n,j}.
\end{equation}
Hence, $s^*_{n,j} = 0$ for $j > n$. The following values of $s^+_{n,j}$ are important for the calculation
of the t.e.v.s in Eqs.~\eqref{eq:tevs_comp}:
\begin{align}
 s^+_{j,j} =& 1, \\
 s^+_{j+1,j} =& \frac{1}{24} (2j+1)(2j+2)(2j+3),\\
 s^+_{j+2,j} =& \frac{1}{5760} (2j+1)(2j+2)(2j+3)(2j+4)(2j+5)(10j+3).
\end{align}

\paragraph{The integral with respect to $k$}
Following the steps in the previous paragraph, the sum over $m$ involving the Bessel functions
in $J_m^*$ is replaced by a sum over $j$ involving powers of $q$. The integral over $k$ can be computed using:
\begin{equation}
 \int_0^p dk\, q^\nu = \frac{\Gamma(\frac{\nu}{2} + 1) \sqrt{\pi}}{2\Gamma(\frac{\nu + 1}{2} + 1)} p^{\nu+1}.
\end{equation}

\paragraph{Analytic expressions in the massless case}
Although the calculation of each individual function $S^*_{abc}$ in Eqs.~\eqref{eq:tevs_comp} has its own
peculiarities, the method is very much the same. To illustrate the method,
the simplest case $a = b = c = 0$ shall be considered for the remainder of this section.

After performing the above steps, $S^+_{000}$ can be brought to the following form:
\begin{multline}\label{eq:sp000_maux}
 S^+_{000} = \frac{1}{\pi^2} \sum_{j= 0}^\infty \frac{(\rho \Omega)^{2j}}{j + \frac{1}{2}} \sum_{n = 0}^\infty
 \frac{\Omega^{2n} s^+_{n + j,j}}{(2n + 2j)!}\\
 \times \int_\mu^\infty dE\, p^{2j + 1}
 \frac{d^{2n+2j}}{dE^{2n+2j}}\left(\frac{1}{1 + e^{\beta E}}\right).
\end{multline}
In the above, the sums over $j$ and $n$ have been swapped, after which the sum over $n$ was shifted down, i.e.
$\sum_{n = 0}^\infty \sum_{j = 0}^n f_{n,j} \rightarrow \sum_{j = 0}^\infty \sum_{n = 0}^\infty f_{n+j,j}$.
While we do not have a method to compute the integral over $E$ in Eq.~\eqref{eq:sp000_maux} for arbitrary
values of the mass $\mu$, in the massless case, $p = E$ and the integration can be done by parts:
\begin{equation}
 \int_0^\infty dE\, E^{2j + 1} \frac{d^{2n+2j}}{dE^{2n + 2j}}\left(\frac{1}{1 + e^{\beta E}}\right) =
 (2j + 1)! \begin{cases}
  \frac{\pi^2}{12\beta^2} & n = 0,\\
  \frac{1}{2} & n = 1, \\
  0 & n > 1.
 \end{cases}
\end{equation}
Then $S^+_{000}$ takes the form:
\begin{equation}
 S^+_{000} = \sum_{j= 0}^\infty (\rho \Omega)^{2j}\left[
 \frac{1}{6\beta^2} + \frac{\Omega^2}{24\pi^2}(2j+3)\right].
\end{equation}
The sum over $j$ can be evaluated as a geometric series, giving:
\begin{equation}
 S^+_{000} = \frac{1}{6\beta^2 \varepsilon} + \frac{\Omega^2}{8\pi^2 \varepsilon^2}
 \left(\frac{2}{3} + \frac{\varepsilon}{3}\right),
\end{equation}
where $\varepsilon = 1 - \rho^2 \Omega^2$.

The above algorithm can be applied to all other terms required for Eqs.~\eqref{eq:tevs_comp}. For brevity,
the individual results shall not be included here, since the terms $S^*_{abc}$ can be inferred from
the final results \eqref{eq:tevs}, together with the relation
\begin{equation}
 S^\times_{110} = \frac{1}{\rho} S^+_{101} - \frac{1}{2\rho} \frac{d}{d\rho} (\rho S^-_{100}).
\end{equation}

\end{document}